\documentclass[prresearch,aps,reprint,longbibliography,superscriptaddress]{revtex4-2}
\pdfoutput=1
\usepackage[utf8]{inputenc}
\usepackage[T1]{fontenc}
\usepackage[dvipsnames]{xcolor}
\usepackage{amsmath}
\usepackage{graphicx}
\usepackage{subfig}
\usepackage{verbatim}
\usepackage{float}
\usepackage[hidelinks]{hyperref}
\usepackage{siunitx}
\usepackage[english]{babel}

\hypersetup{
	colorlinks=true,
	linkcolor=blue,
	urlcolor=blue,
	citecolor=blue,
	pdftitle={Size Dependence in Flux-Flow Hall Effect using Time Dependent Ginzburg-Landau Equations},
	pdfauthor={Vineet Punyamoorty},
}

\usepackage{natbib}

\begin{document}
	\title{Size Dependence in Flux-Flow Hall Effect using Time Dependent Ginzburg-Landau Equations}
	\author{Vineet Punyamoorty}
	\affiliation{Department of Electrical Engineering, Indian Institute of Technology Bombay, Mumbai 400076, India}
	\author{Aditya Malusare}
	\affiliation{Department of Physics and Astronomy, Purdue University, West Lafayette, IN 47907}
	\author{Shamashis Sengupta}
	\affiliation{Universit\'{e} Paris-Saclay, CNRS/IN2P3, IJCLab, 91405 Orsay, France}
	\author{Sumiran Pujari}
	\affiliation{Department of Physics, Indian Institute of Technology Bombay, Mumbai 400076, India}
	\author{Kasturi Saha}
	\affiliation{Department of Electrical Engineering, Indian Institute of Technology Bombay, Mumbai 400076, India}
	\email{kasturis@ee.iitb.ac.in}
	\date{2 February 2021}
	\begin{abstract}
		We study the Hall effect in square, planar type-II superconductors using numerical simulations of time dependent Ginzburg-Landau (TDGL) equations. The Hall field in some type-II superconductors displays sign-change behavior at some magnetic fields due to the induced field of vortex flow, when its contribution is strong enough to reverse the field direction. In this work, we use modified TDGL equations which couple an externally applied current, and also incorporate normal-state and flux-flow Hall effects. We obtain the profile of Hall angle as a function of applied magnetic field for four different sizes $(l\times l)$ of the superconductor: $l/\xi \in \{3,5,15,20\}$. We obtain vastly different profiles for each size, proving that size is an important parameter that determines Hall behavior. We find that electric field dynamics provides an insight into several anomalous features including sign-change of Hall angle, and leads us to the precise transient behavior of order parameter responsible for them.
		
	\end{abstract}
	\maketitle
	
	\section{\label{sec:introduction}Introduction}
	One of the most interesting aspects of superconducting systems is the physics of vortices. Advances in experimental methods have made it possible to probe type-II superconductors at small lengthscales where the behaviour of individual vortices becomes visible. Vortices have been imaged with electron spins in diamonds \cite{nv_prappl, nv_natnano}, scanning superconducting quantum-interference devices (SQUIDs) \cite{squid_nanolett, squid_natcomm} and Hall-probe magnetometry \cite{hall_prb, hall_natmater}. The behaviour of a superconducting system is expected to vary significantly if its dimensions are reduced to be comparable to the coherence length. In this work, we theoretically investigate the effect of finite size on the properties of vortices with numerical simulations of solutions of the time dependent Ginzburg-Landau (TDGL) equations. The flux-flow Hall effect in square planar type-II superconductors are studied for different sample sizes, given by $l/\xi$ ratio of 3, 5, 15 and 20 (where $l$ is the length of the square and $\xi$ is the superconducting coherence length). It will be seen that the electric field and hall angle profiles under a unidirectional current are widely different depending upon the dimension of the superconductor.
	
	Hall effect in superconductors has been observed to display anomalous properties below the critical temperature, most significant of which is the sign-reversal of Hall voltage at certain magnetic fields \cite{artemenko_vortex_1989,iye_hall_1989,hagen_anomalous_1990}. This could not be explained by either the phenomenological models of vortex flow of the time, namely the Bardeen-Stephen (BS) model \cite{bardeen_theory_1965} and the Nozières-Vinen (NV) model \cite{nozieres_motion_1966}, or by the microscopic theory \cite{dorsey_vortex_1992}. It was later proposed that the induced electric field of magnetic vortex flow could contribute to the Hall field and cause anomalous behavior \cite{hagen_anomalous_1990}. Dorsey \cite{dorsey_vortex_1992} and Kopnin \textit{et~al.} \cite{kopnin_flux-flow_1993} proved using analytical approximations of a modified time dependent Ginzburg-Landau (TDGL) system, that indeed sign reversal of Hall effect is possible under some circumstances. These theories make use of microscopic quantities (related to the electronic structure) to define the regimes of sign-reversal. Alternatively, one may numerically compute the Hall effect in a superconducting sample governed by the modified TDGL system of Dorsey \cite{dorsey_vortex_1992} and Kopnin \textit{et~al.} \cite{kopnin_flux-flow_1993}, and find the magnetic field regimes of sign-reversal. This could provide insights into Hall effect behaviour of a superconductor, as a function of macroscopic quantities alone (e.g. GL parameter $\kappa$, sample size, etc.). 
	
	In this work, we follow the alternative numerical route mentioned above, which is direct and does not resort to any analytical approximations. We first use standard time dependent Ginzburg-Landau (TDGL) equations \cite{du_high-kappa_1996,du_numerical_2005,gorkov_vortex_1975} to numerically compute the time-varying order parameter of a planar superconductor in the vortex state. We benchmark our simulations by comparing the numerically obtained fluxoid value of each vortex against $\Phi_o$, the superconducting flux quantum, as a rigorous test (\ref{subsec:vortex_state_and_verification}). With these benchmarks in hand, we simulate a modified TDGL system that includes an externally applied current to probe the Hall effect in these systems. Importantly, we go beyond existing literature by incorporating the normal-state Hall conductivity and flux-flow terms into the dynamical equations that we numerically simulate. In this, we have taken inspiration from the analytical works of Dorsey \cite{dorsey_vortex_1992} and Kopnin \textit{et~al.} \cite{kopnin_flux-flow_1993} (\ref{subsec:flux_flow_under_applied_current}). Next, in \ref{subsec:analysis_of_electric_fields_and_hall_angle} we study the resultant changes in flux-flow and resultant induced electric fields based on the numerical simulation of our modified TDGL system. We compute the Hall angle profile for various sizes of the superconductor, and find vastly different profiles. We find that transient electric fields and related order parameter behavior give us good insight into explaining the anomalous Hall behavior.
	
	\section{\label{sec:theoretical_model}Theoretical Model}
	The TDGL equations are \cite{gorkov_vortex_1975,du_high-kappa_1996,du_numerical_2005}:
	\begin{subequations}
		\begin{multline}
		\gamma\left(\hbar\frac{\partial\psi}{\partial t}+i e_s\Phi\psi\right) + \frac{1}{2m_s}\left(i\hbar\nabla +\frac{e_s}{c}\mathbf{A}\right)^2\psi + \alpha\psi \\ +\beta|\psi|^2\psi = 0
		\end{multline}
		\begin{multline}
		\nu\left(\frac{1}{c}\frac{\partial\mathbf{A}}{\partial t}+\nabla\Phi\right) + \frac{i e_s\hbar}{2m_s}\left(\psi^*\nabla\psi - \psi\nabla\psi^*\right)+\frac{e_s^2}{m_s c}|\psi|^2\mathbf{A} \\ +\frac{c}{4\pi}\nabla\times\nabla\times\mathbf{A}=0    
		\end{multline}
		\label{eqn:tdgl_full}
	\end{subequations}
	where, $\psi$ is the complex-valued order parameter, $\alpha$ and $\beta$ are the phenomenological parameters of Ginzburg-Landau theory \cite{landau_superconductivity_1950,tinkham_introduction_1996} and $\phi$, $\mathbf{A}$ are the electric and magnetic potentials respectively. $\nu$ is the normal-state conductivity and $\gamma$ is a relaxation constant for the order parameter. The charge and mass of Cooper pairs are denoted by $e_s = 2e$ and $m_s = 2m_e$ respectively. 
	\subsection{\label{subsec:gauge_invariance_and_normalization}Gauge invariance and normalization}
	For an arbitrary function $\chi(x,y,t)$ with well-defined spatial and time derivatives, the gauge invariance can be expressed as ($\psi \rightarrow \psi e^{i\kappa\chi}$, $\mathbf{A} \rightarrow \mathbf{A}+\nabla\chi$, $\phi \rightarrow \phi-\frac{\partial\chi}{\partial t}$) . We choose the zero electric potential gauge (i.e. $\phi=0$) \cite{du_numerical_2005,deang_study_1997,du_high-kappa_1996}. The physical quantities in \eqref{eqn:tdgl_full} are renormalized as follows:
	\begin{align}
	\frac{x}{\xi}\rightarrow x,\hspace{0.8em}\frac{t}{\gamma\hbar/(-\alpha)}\rightarrow t, \hspace{0.8em}
	\frac{\mathbf{A}}{\sqrt{2}H_c\xi} \rightarrow \mathbf{A}, \hspace{0.8em}\frac{\psi}{\sqrt{-\alpha/\beta}} \rightarrow \psi
	\label{eqn:normalization}
	\end{align}
	where $H_c=\sqrt{\frac{4\pi\alpha^2}{\beta}}$ and $\xi$ is the GL coherence length \cite{deang_study_1997}. In the chosen gauge, the resulting dimensionless equations applicable over the superconducting domain are:
	\begin{subequations}
		\label{eqn:tdgl_final}
		\begin{equation}
		\frac{\partial\psi}{\partial t}+\left(i\nabla+\frac{1}{\kappa}\mathbf{A}\right)^2\psi-\left(1-|\psi|^2\right)\psi = 0 
		\label{eqn:tdgl_final_1}
		\end{equation}
		\begin{multline}
		\sigma\frac{\partial\mathbf{A}}{\partial t}+\nabla\times\nabla\times\mathbf{A}+\frac{i}{2\kappa}\left(\psi^*\nabla\psi-\psi\nabla\psi^*\right) + \\ \frac{1}{\kappa^2}|\psi|^2\mathbf{A} = 0
		\label{eqn:tdgl_final_2}
		\end{multline}
	\end{subequations}
	where, $\kappa$ is the GL parameter and $\sigma$ is normalized normal-state conductivity. When the sample is surrounded by vacuum on all sides and no external current is passed, the following boundary conditions (BCs) apply \cite{ogren_self-consistent_2012,du_high-kappa_1996,deang_study_1997,alstrom_magnetic_2011}:
	\begin{align}
	\label{eqn:bc_vacuum_sc}
	\nabla\times\mathbf{A}&=\mathbf{H}_\text{ext},&
	\left(i\nabla+\frac{1}{\kappa}\mathbf{A}\right)\psi\cdot\mathbf{n}&=0,& 
	-\sigma\frac{\partial\mathbf{A}}{\partial t}\cdot\mathbf{n}&=0
	\end{align}
	(where $\mathbf{H}_{\text{ext}}$ is the externally applied magnetic field, perpendicular to the sample). The first condition imposes continuity of transverse magnetic field across the boundary, while the second and third ensure that neither supercurrent nor normal current crosses the boundary, respectively. 
	
	In principle, one also needs to solve Maxwell's equation $\nabla\times\nabla\times\mathbf{A} = 0$ (considering vacuum) in the surrounding domain \cite{deang_study_1997}. However, in the case of two-dimensional samples (i.e. infinitely long cylinders) with perpendicularly applied magnetic field, as in our system, solving the ``interior problem'' \eqref{eqn:tdgl_final} alone is sufficient to a good approximation \cite{du_numerical_2005,chapman_simplified_1995}. For reference, the ``full problem'' and its boundary conditions are discussed in Refs.~\cite{du_high-kappa_1996} and \cite{deang_study_1997}.
	
	\subsection{\label{subsec:inclusion_of_extrenally_applied_current}Inclusion of externally applied current}
\begin{figure}
		\centering
		\includegraphics[width=\columnwidth]{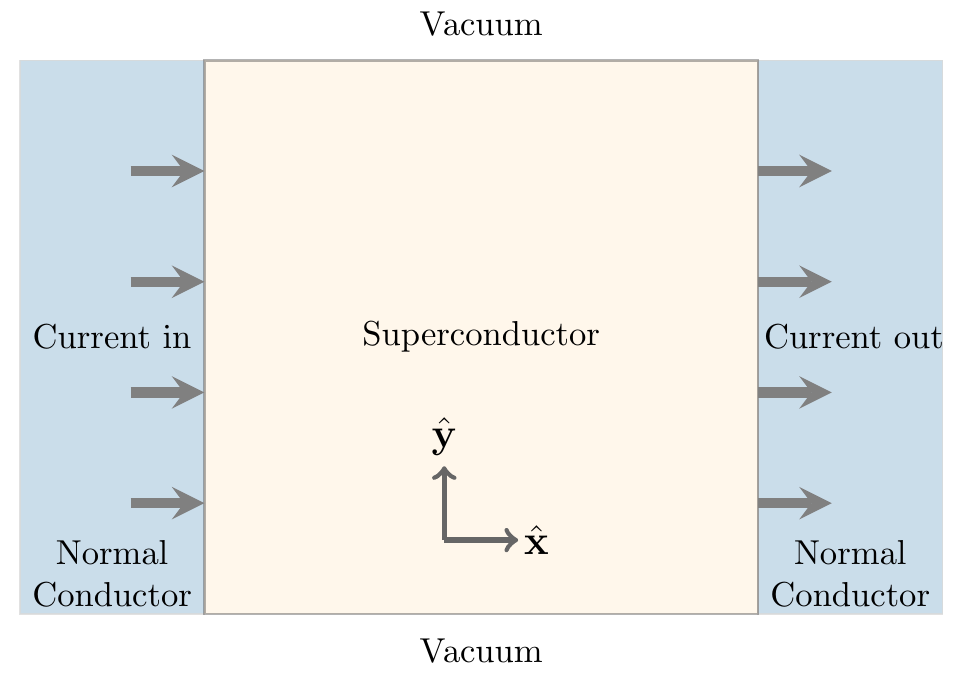}
		\caption{Schematic representation of our simulation domain when an external current is applied -- $l\times l$ square superconductor with metallic contacts on the left and right, and vacuum at top and bottom. Current $\mathbf{J}_a$ is applied along $\hat{\mathbf{x}}$ and magnetic field $\mathbf{H}_\text{ext}$ along $\hat{\mathbf{z}}$.}
		\label{fig:domain}
	\end{figure}

	Including an externally applied transport current in the TDGL system entails two tasks: (a) accounting for the magnetic field induced by the transport current and (b) modifying the boundary conditions \eqref{eqn:bc_vacuum_sc} to account for flow of normal current across boundaries. The former can be achieved by modifying the first boundary condition to $\nabla\times\mathbf{A} = \mathbf{H}_{\text{tot}}$ where $\mathbf{H_\text{tot}} = \mathbf{H}_\text{ext}+\mathbf{H}_c$ is the sum of applied ($\mathbf{H}_\text{ext}$) and induced ($\mathbf{H}_c$) magnetic fields \cite{ogren_self-consistent_2012}. $\mathbf{H}_c$ is to be computed from the current profile (a function of $\psi$, $\mathbf{A}$), making the system self-consistent. However, an approximation is frequently used in literature \cite{machida_direct_1993,gropp_numerical_1996,winiecki_time-dependent_2002,vodolazov_dynamics_2004,ogren_self-consistent_2012} to simplify the computation of $\mathbf{H}_c$: the current profile is assumed to be a uniform band, which reduces $\mathbf{H}_c$ to a simple expression involving $\mathbf{J}_a$, the applied (uniform) current density. In this paper, $\mathbf{J}_a$ is assumed to be in the $+\mathbf{\hat{x}}$ direction (fig.~\ref{fig:domain}), which gives us $\mathbf{H}_c$ along all four boundaries as $\mathbf{H}_{c}^\text{top, bottom} = \pm WJ_a/2~\mathbf{\hat{z}}$ (where $W$ is the length of the sample along $\mathbf{\hat{y}}$) and $\mathbf{H}_{c}^\text{left, right}$ vary linearly between the bottom and top edges.
	
	Secondly, to account for flow of normal current across the boundary, the ``vacuum-superconductor'' BCs \eqref{eqn:bc_vacuum_sc} are now replaced by ``metal-superconductor'' BCs \eqref{eqn:bc_metal_sc_tot} \cite{vodolazov_dynamics_2004,ogren_self-consistent_2012} at the left and right edges, while retaining the former BCs \eqref{eqn:bc_vacuum_sc_tot} at the top and bottom edges:
	\begin{subequations}
		\label{eqn:bc_tot}
		\begin{align}
		\label{eqn:bc_vacuum_sc_tot}	
		&\nabla\times\mathbf{A}=\mathbf{H}_\text{tot},\quad
		\left(i\nabla+\frac{1}{\kappa}\mathbf{A}\right)\psi\cdot\mathbf{n}=0,\quad -\sigma\frac{\partial\mathbf{A}}{\partial t}\cdot\mathbf{n}=0\\
		\label{eqn:bc_metal_sc_tot}
		&\nabla\times\mathbf{A}=\mathbf{H}_\text{tot},\qquad\psi=0,\qquad 
		-\sigma\frac{\partial\mathbf{A}}{\partial t}\cdot\mathbf{n}=\mathbf{J}_a\cdot\mathbf{n}
		\end{align}
	\end{subequations}
	In \eqref{eqn:bc_metal_sc_tot}, the third BC accounts for flow of current across the boundary, and the second condition ensures that the density of superconducting electrons is zero at the edges. This ensures that the injected normal current transitions to supercurrent gradually, rather than abruptly, inside the superconductor.
	
	\subsection{\label{subsec:inclusion_of_hall_effect}Inclusion of Hall effect}
	Normal state hall effect is a result of the conduction of electrons transverse to applied electric field. This can be incorporated in the TDGL system \eqref{eqn:tdgl_final}, \eqref{eqn:bc_tot} by rewriting the normal-state conductivity $\sigma$ as a tensor:
	\begin{equation}
	\sigma = \begin{pmatrix}
	\sigma_{xx} & \sigma_{xy} \\
	\sigma_{yx} & \sigma_{yy}
	\end{pmatrix}
	\end{equation}
	We assume an isotropic sample ($\sigma_{xx} = \sigma_{yy}$), and by symmetry $\sigma_{yx} = -\sigma_{xy}$. In order to determine $\sigma_{xx}$ and $\sigma_{xy}$, we use the following as the model for the normal-state conductivity \cite{hurd_hall_1972,dorsey_vortex_1992}:
	\begin{align}
	\sigma_{xx} &= \sigma_0\frac{1}{1+\omega_c^2\tau^2}, &\sigma_{xy} &= \sigma_0\frac{\omega_c\tau}{1+\omega_c^2\tau^2} \label{eqn:sigma_twoband_full}
	\end{align}
	where $\omega_c$ is the cyclotron frequency $e_s\left(\nabla\times\mathbf{A}\right)/m_s$ and $\tau$ is the electron scattering time. Under typical conditions, we have $\omega_c\tau \ll 1$ (low-field limit) \cite{hurd_hall_1972,dorsey_vortex_1992}. Thus, $\sigma_{xx} \approx \sigma_0$ and $\sigma_{xy} \approx \omega_c\tau\sigma_0$. Due to the spatially varying magnetic field, cyclotron frequency $\omega_c$ and consequently, $\sigma_{xy}$ are also spatially varying. In order to enforce the low-field limit, we take $\omega_c\tau = 10^{-2}\left(\nabla\times\mathbf{A}\right)$, where the pre-factor of $10^{-2}$ ensures that $\omega_c\tau \ll 1$.
	
	Josephson \cite{josephson_potential_1965} proved that macroscopically, a vortex moving at velocity $\mathbf{v}_L$ gives rise to an induced electric field $\mathbf{E} = -\frac{1}{c}\mathbf{v}_L\times\mathbf{H}$,where all the quantities are spatially and temporally averaged \cite{hagen_anomalous_1990,dorsey_vortex_1992,kopnin_flux-flow_1993}. In the Bardeen-Stephen (BS) model \cite{bardeen_theory_1965}, under an applied current $\mathbf{J}_a$, vortices experience a Lorentz force $\sim \mathbf{J}_a\times\mathbf{H}$ \cite{dorsey_vortex_1992,hagen_anomalous_1990}. This force gives rise to velocity $\mathbf{v}_L$ along $\mathbf{J}_a\times\mathbf{H}$ and in a system such as ours (fig.~\ref{fig:domain}), this results in vortex motion along $-\hat{\mathbf{y}}$. This velocity $\mathbf{v}_L$ along $-\hat{\mathbf{y}}$ produces a field $\mathbf{E} = -\frac{1}{c}\mathbf{v}_L\times\mathbf{H}$ along $\hat{\mathbf{x}}$, the same direction as $\mathbf{J}_a$, thereby causing dissipation. However, if we were to have an additional component of $\mathbf{v}_L$ along $\hat{\mathbf{x}}$ (the same direction as $\mathbf{J}_a$), this would create a field contribution in the Hall direction.
	Dorsey \cite{dorsey_vortex_1992} and Kopnin \textit{et al.} \cite{kopnin_flux-flow_1993} proved that adding a non-zero imaginary part to the relaxation parameter $\gamma$ in the TDGL system \eqref{eqn:tdgl_full} produces such flux-flow contribution to the Hall field by lending the vortices a velocity component parallel to the applied current. Thus, we write $\gamma = \gamma_1+i\gamma_2$. In the microscopic picture, the value of $\gamma_2/\gamma_1$ has been shown to depend upon the electronic structure of the material \cite{dorsey_vortex_1992,kopnin_flux-flow_1993}. The sign of $\gamma_2/\gamma_1$ determines whether vortices travel along $\mathbf{J}_a$ or against, and therefore crucially affects the sign of flux-flow contribution to Hall field. Kopnin \textit{et al.} \cite{kopnin_flux-flow_1993} proved that a sign reversal in Hall effect would be observed for negative values of $\gamma_2/\gamma_1$.
	
	\section{Results and discussion}
	We use COMSOL Multiphysics\textsuperscript{\textregistered} \cite{comsol}, a commercial finite element tool to numerically simulate TDGL equations \eqref{eqn:tdgl_final}--\eqref{eqn:bc_tot}. Throughout the paper, we take GL parameter $\kappa=2$ and normalized normal-state conductivity $\sigma_0=1$. In section~\ref{subsec:vortex_state_and_verification}, we obtain a vortex state solution and propose a procedure to rigorously verify the solution to help identify any numerical errors or artifacts. In sections \ref{subsec:flux_flow_under_applied_current}--\ref{subsec:analysis_of_electric_fields_and_hall_angle}, we study a system with externally applied current and Hall effect included. Throughout these simulations, we apply an external current $\mathbf{J}_a = 0.04\thinspace\hat{\mathbf{x}}$. We consider four geometries: $l\times l$ square superconductors for $l/\xi\in\{3,5,15,20\}$. We apply an external magnetic field $\mathbf{H}_\text{ext}$ along $\hat{\mathbf{z}}$, whose value is swept between 0 and $H_{c2} = \kappa = 2$ in steps of 0.05. For each combination of size $l$ and $\mathbf{H}_\text{ext}$, we solve the modified TDGL system \eqref{eqn:tdgl_final}, \eqref{eqn:bc_tot} and compute the time-varying order parameter $\psi\left(\mathbf{r},t\right)$ and vector potential $\mathbf{A}\left(\mathbf{r}, t\right)$. This gives us a complete insight into the dynamic vortex motion and electric fields $E_x = -\partial{A}_x/\partial t$ and $E_y = -\partial{A}_y/\partial t$, which form the basis for much of our analysis. 
	
	Since we use normalized units throughout the paper, we compute the order of magnitude for these units to understand the typical physical values. We chose $\kappa = 2$, which broadly corresponds to Niobium \cite{finnemore_superconducting_1962}. Using the physical parameters of Nb, we compute the time units as $\sim \SI{e-12}{\second}$ \cite{oripov_time-dependent_2020}, current density units as $\sim\SI{1}{\ampere\per\metre}$ \cite{asada_superconductivity_1969,williamson_nonlocal_1970} and the applied magnetic field units as $\sim1$\thinspace{kOe} \cite{williamson_nonlocal_1970}.
	
	\subsection{\label{subsec:vortex_state_and_verification}Vortex state and verification}
\begin{figure}
		\centering
		\includegraphics[width=\columnwidth]{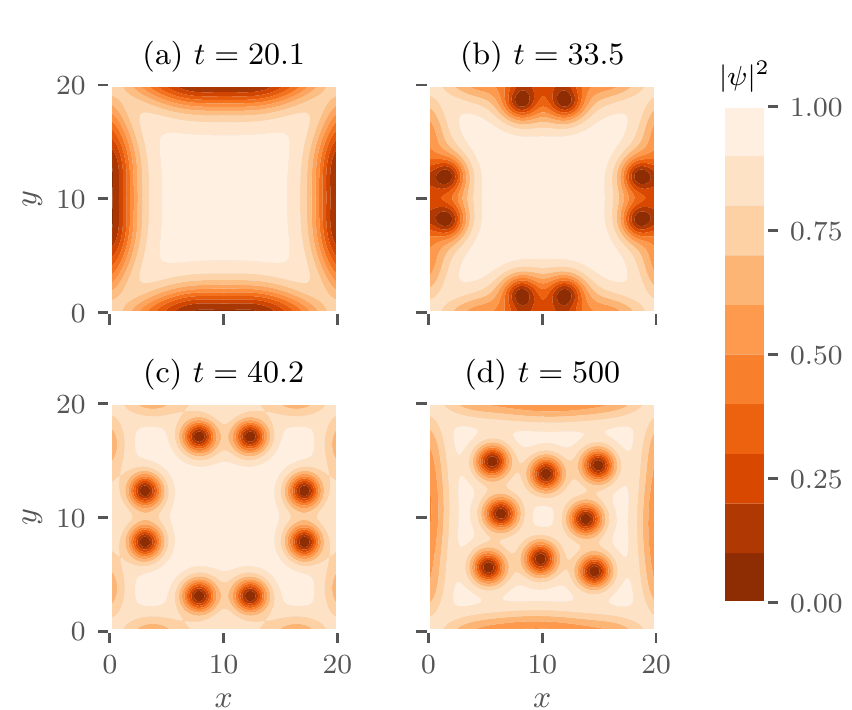}
		\caption{$|\psi|^2$ at various instants of time, depicting major events in the formation of a vortex state in the $20\xi\times20\xi$ sample. Applied magnetic field is $|\mathbf{H}_\text{ext}| = 0.9$.}
		\label{fig:vortex_state}
	\end{figure}

\begin{figure}
		\centering
		\includegraphics[scale=1]{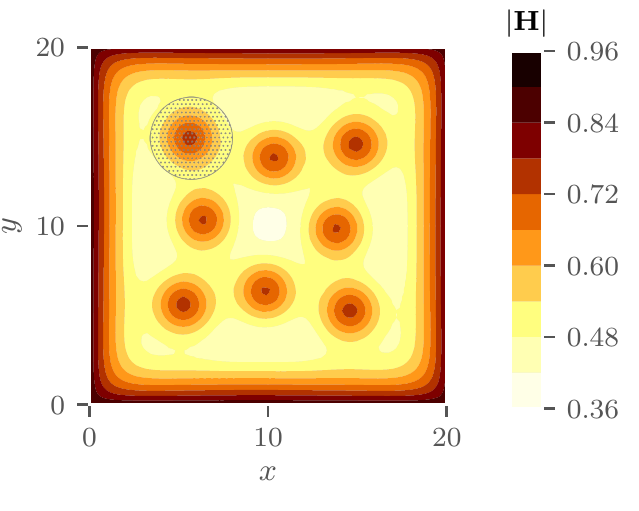}
		\caption{We verify that the fluxoid value of each of our vortices is  $\approx \Phi_0$. To compute the fluxoid value we take a circular area centred at the vortex-centre defined by the local maxima in magnetic field, with a radius $2.4\xi$ (shown with a dotted hatch on the top-left vortex). Figure shown for $|\mathbf{H}_\text{ext}|=0.9$.}
		\label{fig:k2b0.9_verif}
	\end{figure}

	We first solve TDGL equations \eqref{eqn:tdgl_final} on a $20\xi\times20\xi$ planar sample (GL parameter $\kappa=2$) with no applied current or Hall effect enabled. Thus, in this case we use the vacuum-superconductor BCs \eqref{eqn:bc_vacuum_sc} on all four sides, and observe a vortex state solution at $\mathbf{H}_\text{ext} = 0.9\hat{\mathbf{z}}$ (fig.~\ref{fig:vortex_state}).
	We perform a thorough verification of our simulation as described in the following. First, we confirm that the simulation results are stable with respect to mesh size. We then sweep $|\mathbf{H}_\text{ext}|$ widely and observe the existence of both upper and lower critical fields marked by the vanishing of vortices, as expected from a type-II superconductor ($\kappa>{1}/{\sqrt{2}}$). We also observe that for the same $\kappa$, the number of vortices increases (decreases) when the external field $|\mathbf{H}_\text{ext}|$ is increased (decreased). This is an expected qualitative behaviour of the superconductor to let more (less) incident flux pass through the superconductor. One of the hallmarks of superconducting vortices under Ginzburg-Landau theory is \textit{fluxoid} quantization. It states that the fluxoid value $\Phi'$ associated with each vortex is quantized by $\Phi_0 = \frac{hc}{2e}$ \cite{tinkham_introduction_1996}.
	\begin{equation}
	\Phi' = \Phi + \frac{4\pi}{c}\oint\lambda^2\mathbf{J}_s\cdot dl
	\label{eqn:fluxoid_quantum}
	\end{equation}
	where $\Phi$ is the magnetic flux $\int\mathbf{B}\cdot d\mathbf{s}$ associated with the vortex, and the integral term involving super current $\mathbf{J}_s$ and penetration depth $\lambda$ is performed on a closed contour enclosing the vortex. We compute the fluxoid value, and obtain $\Phi'\approx0.96\Phi_0$ for each vortex, confirming that our simulations firmly uphold fluxoid quantization (fig.~\ref{fig:k2b0.9_verif}).
	
	\subsection{\label{subsec:flux_flow_under_applied_current}Flux flow under applied current with Hall effect incorporated}
\begin{figure}
		\centering
		\includegraphics[width=\columnwidth]{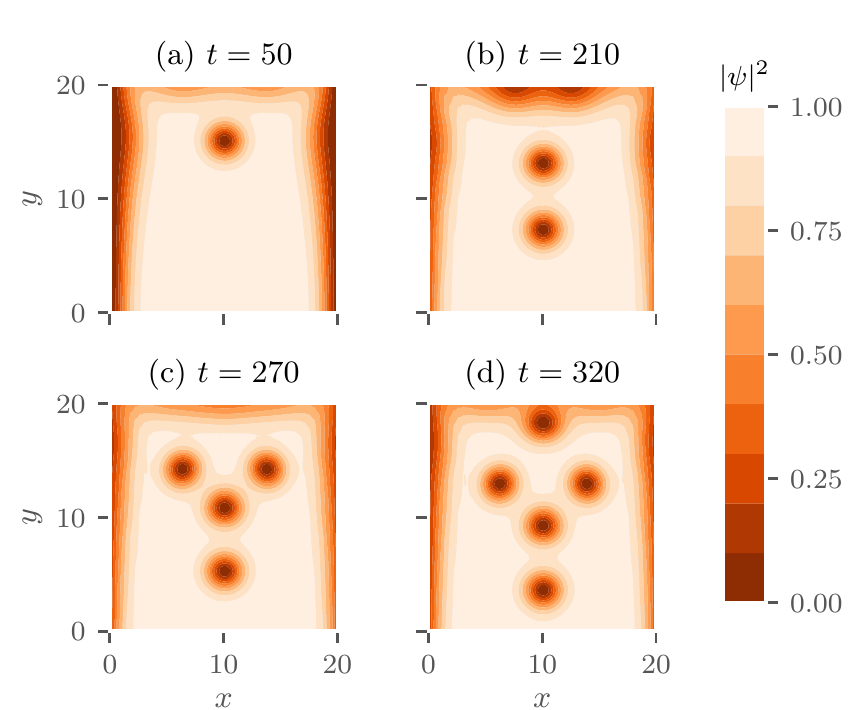}
		\caption{We observe vortex motion under an applied external current in the transverse direction. Vortices enter at the top edge and leave at the bottom. This motion is followed by the superconductor being driven into normal state. ($|\mathbf{H}_\text{ext}| = 0.5$)}
		\label{fig:applied_current}
	\end{figure}

	In the Bardeen-Stephen (BS) model \cite{bardeen_theory_1965}, magnetic vortices experience a Lorentz force under applied external current along $\mathbf{J}_a\times{\mathbf{H}}$ \cite{hagen_anomalous_1990,dorsey_vortex_1992}. We apply an external current $\mathbf{J}_a = 0.04\thinspace\hat{\mathbf{x}}$ and $\mathbf{H}_\text{ext}$ along $+\hat{\mathbf{z}}$, and accordingly observe vortex flow along $-\hat{\mathbf{y}}$ with vortices entering at the top edge and leaving at the bottom (fig.~\ref{fig:applied_current}). Consistent with the BS model, we observe faster (slower) motion of vortices with increased (decreased) magnitudes of $\mathbf{J}_a$. Choosing a larger value of $|\mathbf{J}_a|$ would increase the magnitude of current-induced magnetic field $\mathbf{H}_c$, driving the superconductor into normal state (after some time) for even low values of applied field $\mathbf{H}_\text{ext}$. With our chosen value of $|\mathbf{J}_a| = 0.04$, we find that the system is ultimately driven into normal state for all values of $|\mathbf{H}_\text{ext}| \ge 0.35$ in the case of sizes $l/\xi\in\{15,20\}$. For the other two smaller sizes, the system is driven into normal state for all values of $|\mathbf{H}_\text{ext}|$.
	
	For sizes $l/\xi\in\{15,20\}$, we observe vortices and their movement, before the system eventually goes into normal state ($\psi\left(\mathbf{r}\right) = 0$), whereas for sizes $l/\xi\in\{3,5\}$, we do not observe vortices for any value of $|\mathbf{H}_\text{ext}|$. To understand further, we solve the standard TDGL system \eqref{eqn:tdgl_final}, \eqref{eqn:bc_vacuum_sc} (with no applied current) for $l/\xi\in\{3,5\}$ and sweep the magnetic field $|\mathbf{H}_\text{ext}|$ as earlier. We do not find a vortex state solution for any value of $|\mathbf{H}_\text{ext}|$. We conclude that this is a size-effect: the system is smaller than a critical size, forbidding the possibility of a vortex state solution.
	
\begin{figure}
		\centering
		\includegraphics[width=\columnwidth]{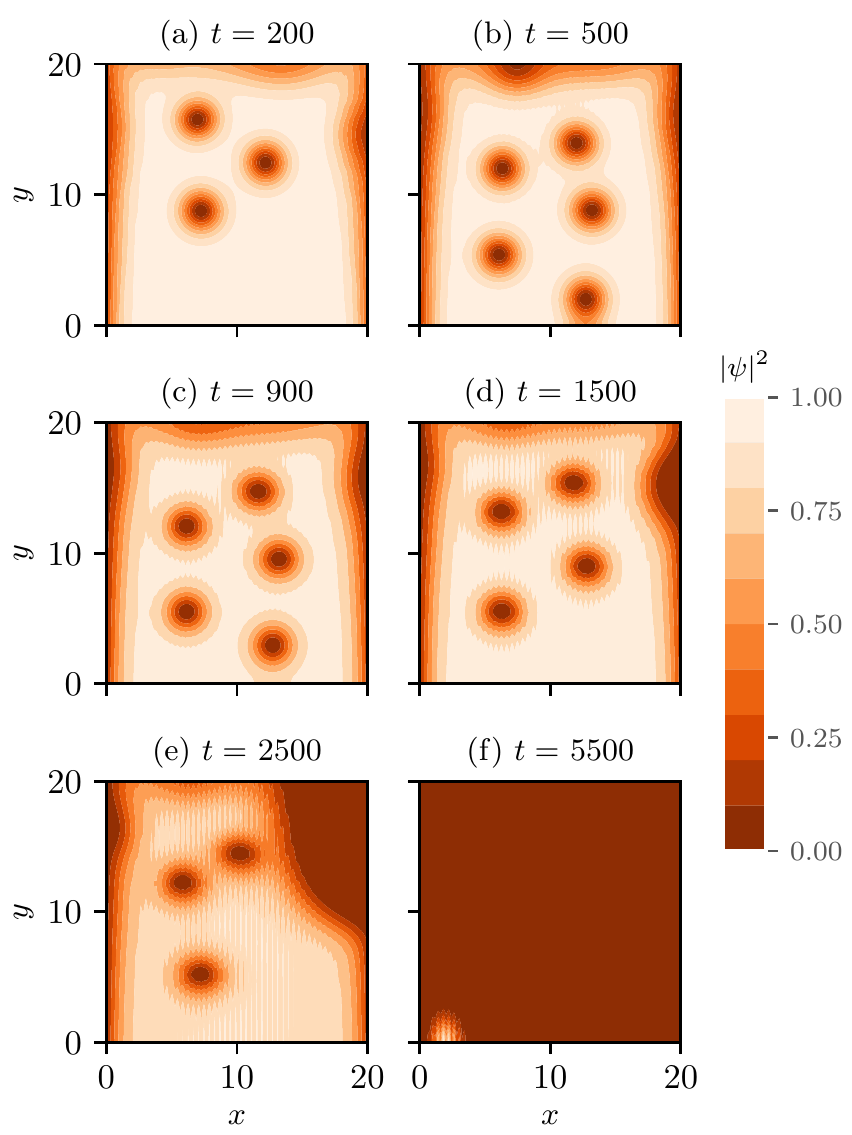}
		\caption{Complex vortex flow under an applied external current with Hall effect enabled. The setting in of normal state from the right edge starting at $t=1500$ induces a gradual transient field largely affecting Hall behavior (fig.~\ref{fig:efield2x1}). ($|\mathbf{H}_\text{ext}|=0.5$ and $\gamma_2/\gamma_1 = 0.4$)}
		\label{fig:ordPar3x2}
	\end{figure}

	Next, we also enable normal-state and flux-flow Hall effects following the discussion in \ref{subsec:inclusion_of_hall_effect}, with $\gamma_2/\gamma_1 = 0.4$ and $-0.4$ as separate cases. The choice of $|\gamma_2/\gamma_1|$ was made in such a way that the vortices gain a perceivable amount of velocity in the direction of $\mathbf{J}_a$, but not significantly enough, so that vortices still primarily move along $-\hat{\mathbf{y}}$. With Hall effect enabled, we observe complex vortex motion (fig.~\ref{fig:ordPar3x2}). Vortices enter the system at the top edge and traverse smooth but irregular trajectories through the domain, due to complex vortex-vortex and vortex-boundary interactions. The key difference is that they also obtain velocity in the $\pm\hat{\mathbf{x}}$ direction, unlike when Hall effect is not enabled.
	We seek to capture the effect of this complex motion on the longitudinal and Hall electric fields and explain the observed behavior.
	\subsection{\label{subsec:analysis_of_electric_fields_and_hall_angle}Analysis of electric fields and Hall angle}
	
\begin{figure}
		\subfloat{\includegraphics[width=\columnwidth]{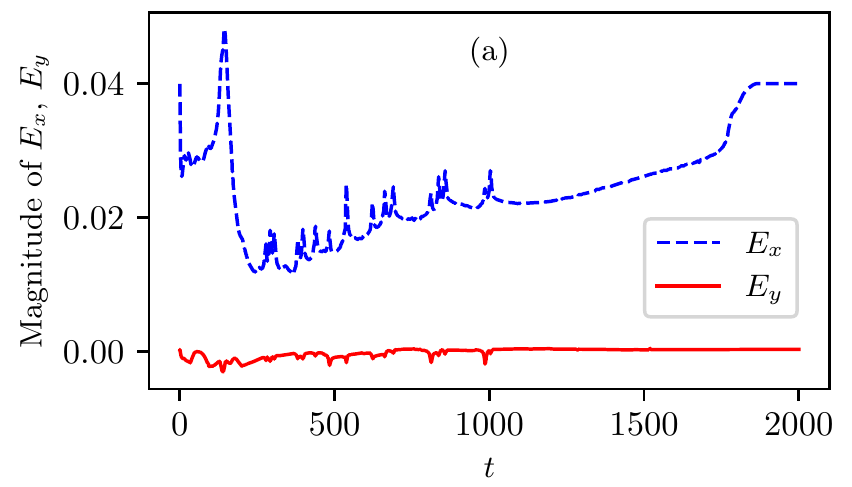}\label{subfig:efield1}}\\
		\subfloat{\includegraphics[width=\columnwidth]{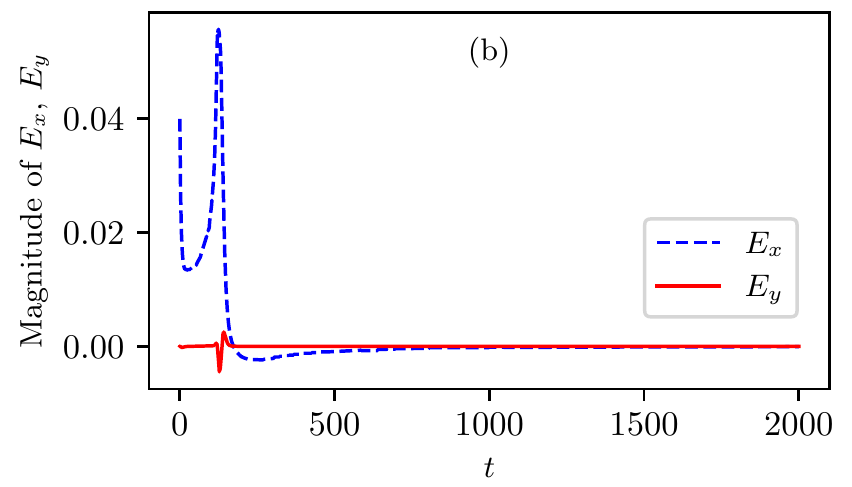}\label{subfig:efield2}}\\
		\subfloat{\includegraphics[width=\columnwidth]{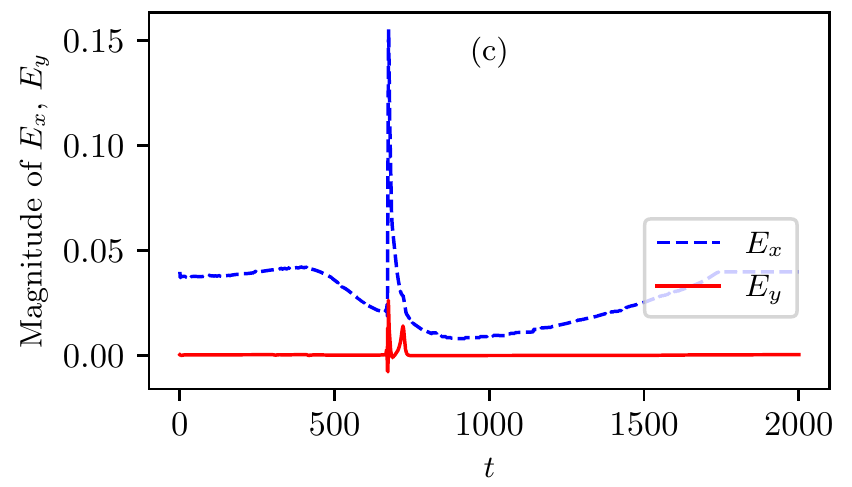}\label{subfig:efield3}}
		\caption{The three different types of electric field profiles observed in our simulations across all sizes and magnitudes of applied field $|\mathbf{H}_\text{ext}|$. (a) Saturates to $\sigma_0J_a$, exhibits spiking behavior corresponding to entry and exit of vortices from sample of size $l/\xi=20$, and $|\mathbf{H}_\text{ext}|=0.8$. (b) Saturates to $0$, of sample size $l/\xi=20$, and $|\mathbf{H}_\text{ext}|=0.2$). (c) Saturates to $\sigma_0J_a$, for sample size $l/\xi=5$, and $|\mathbf{H}_\text{ext}|=1.2$.}
		\label{fig:efields}
	\end{figure}

	We obtain the spatially averaged electric field by averaging across the entire domain.	We find that the so obtained $\mathbf{E}$ profiles can be grouped into three distinct types (fig.~\ref{fig:efields}) of behaviour based on the $l/\xi$ ratio and value of $|\mathbf{H}_\text{ext}|$. First, at a large $l/\xi$ of 20 (fig.~\ref{subfig:efield1}), the longitudinal field saturates to $\sigma_0J_a$ (normal state) with the observation of a series of spikes prior to that. This spiking occurs due to vortex entry and exit, as discussed in Ref. \cite{machida_direct_1993}. The Hall field $E_y$ also exhibits similar spiking as a result of the vortex velocity component in the direction of $\mathbf{J}_a$. Such behaviour occurs for sizes $l/\xi \in\{15,20\}$ and applied field $|\mathbf{H}_\text{ext}|\ge 0.35$. For fields lower than $0.35$, the system rapidly evolves to the superconducting Meissner state with $E_x$ reaching zero as shown in fig.~\ref{subfig:efield2} for $|\mathbf{H}_\text{ext}|$ of 0.2. Sizes $l/\xi \in \{3,5\}$ are always driven into normal state. This results in the third type, with $E_x$ saturating to $\sigma_0J_a$ (fig.\ref{subfig:efield3}). These differences in electric field behavior produce significantly different Hall effect profiles, which we address next.
	
\begin{figure}
		\includegraphics[width=\columnwidth]{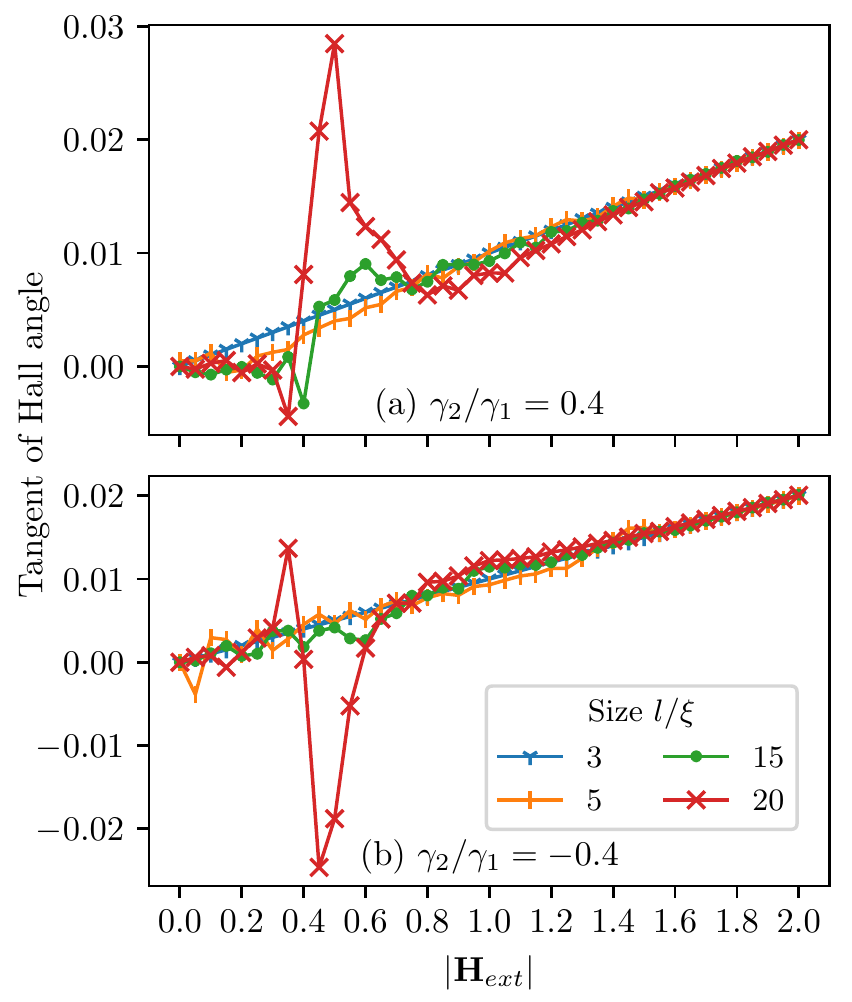}
		\caption{(a)Hall angle profiles for $\gamma_2/\gamma_1=+0.4$.(b) Hall angle profiles for $\gamma_2/\gamma_1=-0.4$. Evidently, size is an important parameter in determining Hall behavior, with the smallest size $l/\xi=3$ behaving identical to a normal conductor. Sign-reversal is only seen for $l/\xi=20$, in which case $\gamma_2/\gamma_1=0.4$ and $-0.4$ show opposite behavior. Fields were averaged from $t = 0$ to $11000$. }
		\label{fig:hall_1}
	\end{figure}

	We characterize the Hall effect using Hall angle, the ratio of effective transverse to longitudinal conductivity: $\tan{\theta_H} = \sigma_{xy}'/\sigma_{xx}'$ \cite{dorsey_vortex_1992,kopnin_flux-flow_1993}. These macroscopic effective conductivties are marked with a prime to distinguish them from the normal-state quantities $\sigma_{xx}$ and $\sigma_{xy}$ \eqref{eqn:sigma_twoband_full}. From the macroscopic equation $\mathbf{J}_a = \sigma'\cdot\mathbf{E}$, we have: 
	\begin{align}
	\begin{pmatrix}
	J_a \\
	0
	\end{pmatrix} &= \begin{pmatrix}
	\sigma_{xx}' & \sigma_{xy}' \\
	-\sigma_{xy}' & \sigma_{xx}'
	\end{pmatrix} \cdot \begin{pmatrix}
	E_x \\
	E_y 
	\end{pmatrix}
	\label{eqn:hall_angle_derv}
	\end{align}
	Thus, we get $\tan{\theta_H} = \sigma_{xy}'/\sigma_{xx}' = E_y / E_x $, where the fields are both spatially and temporally averaged. We obtain the tangent of Hall angle $\tan{\theta_H}$ for each combination of applied magnetic field $|\mathbf{H}_\text{ext}|$ and size $l/\xi \in \{3,5,15,20\}$ (fig.~\ref{fig:hall_1}). We first note that the Hall angle profiles resulting from our choice of $|\gamma_2/\gamma_1|$ closely resemble the experimental data with respect to orders of magnitude ($\tan{\theta_H}\sim 10^{-2}$) \cite{iye_hall_1989,forro_flux-flow_1989}.
	
	We find that as $|\mathbf{H}_\text{ext}| \rightarrow H_{c2}$, when the superconductor is rapidly driven into normal state, Hall angle varies linearly with $|\mathbf{H}_\text{ext}|$, as in the case of normal metals. In this regime, flux-flow contribution is negligible and all sizes $l/\xi \in \{3,5,15,20\}$ have the same profile. This is in agreement with expected behavior because we do not have any size-effect for normal metals. When $|\mathbf{H}_\text{ext}| \approx 0$, the Hall angle approaches 0. At these low fields, sizes $l/\xi \in \{3,5\}$ are driven to normal-state. As normal metals, they have negligible Hall fields at $|\mathbf{H}_\text{ext}| \approx 0$, and consequently a very small Hall angle. On the other hand, sizes $l/\xi \in \{15,20\}$ are in the superconducting state at these low fields, and therefore have negligible  $E_y$ and $E_x$. The small non-zero contribution is a result of transient behavior as seen in fig.~\ref{subfig:efield2}. Thus, although the Hall angle is small for both groups of sizes at $|\mathbf{H}_\text{ext}| \approx 0$, the underlying states are different.
	
	We next note that the Hall angle profile is completely linear for size $l/\xi  = 3$. This is a result of it behaving as a normal metal with no vortex state, and electric field $E_x\left(t\right)$ saturating to $\sigma_0 J_a$ for all values of $|\mathbf{H}_\text{ext}|$ (third type shown in fig.~\ref{subfig:efield3}). However, size $l/\xi=5$ also exhibits identical behavior, but we find significant deviations from linearity. To understand further, we look at the difference in transient fields between the two sizes. Large transients in $E_x\left(t\right)$ are common for size $l/\xi = 5$, such as the one shown in fig.~\ref{subfig:efield3}, whereas the transient is highly suppressed in size $l/\xi = 3$. In fact, across all values of $|\mathbf{H}_\text{ext}|$, the maximum percentage change in $E_x\left(t\right)$ (relative to saturation value $\sigma_0 J_a$) is limited to $\approx 0.02\%$ for size $l/\xi = 3$, and to $\approx 760\%$ in the case of size $l/\xi = 5$. We conclude that this enormous difference in transient levels leads to deviations from linearity in size $l/\xi = 5$, in spite of non-superconducting behavior.
	
\begin{figure}
		\includegraphics[width=\columnwidth]{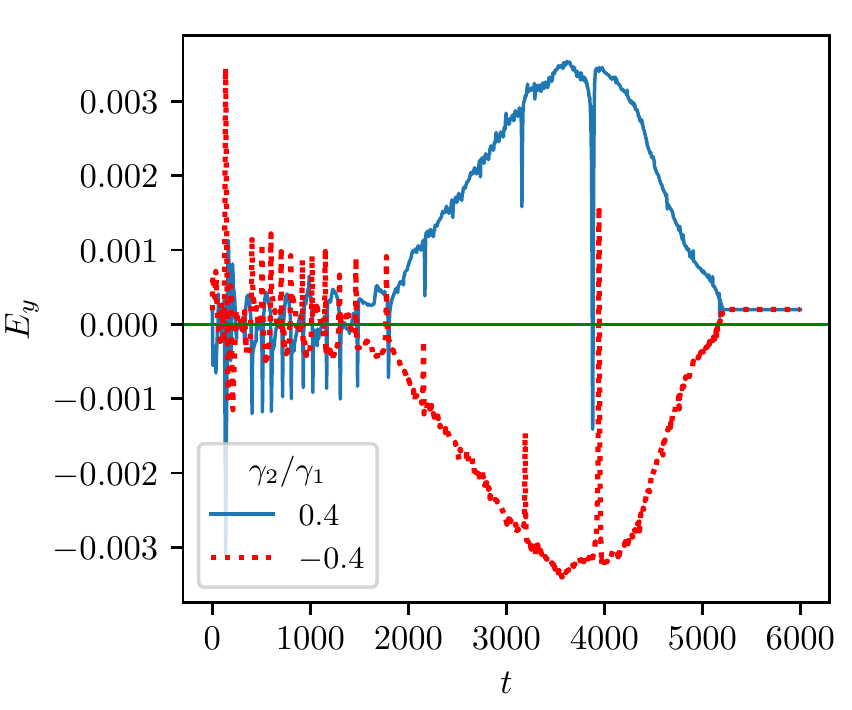}
		\caption{$E_y$ fields for size $l/\xi=20$ at $|\mathbf{H}_\text{ext}|=0.5$} The gradual transients starting at $t\approx 1500$ are responsible for the vastly different Hall angle for $\gamma_2/\gamma_1 = 0.4$ and $-0.4$. Similar transients are seen for $|\mathbf{H}_\text{ext}|\approx0.5$, leading to the Hall angle profile in fig.~\ref{fig:hall_1}.
		\label{fig:efield2x1}
	\end{figure}

	We next look at sign-reversal of Hall angle, a key anomaly in the Hall behavior of some superconductors. For size $l/\xi = 20$, we observe a 
	region of sharp deviation from linear, ``normal-conductor'' behavior for intermediate values of applied field, i.e. around $|\mathbf{H}_\text{ext}| \approx 0.5$ (fig.~\ref{fig:hall_1}). This is the region where flux-flow contribution to the Hall effect is most significant, because for higher fields, normal state sets in rapidly, and for lower fields, there are no vortices (Meissner state).
	We find that strong, negative flux-flow contribution leads to sign-reversal for $\gamma_2/\gamma_1 = -0.4$, and positive contribution leads to a peak for $\gamma_2/\gamma_1 = 0.4$, in agreement with Kopnin \textit{et al.} \cite{kopnin_flux-flow_1993}. In order to understand the precise transient behavior leading to this vast difference, we look at the spatially averaged fields $E_x$ and $E_y$ as functions of time (for $|\mathbf{H}_\text{ext}|=0.5$). First, we find that for both $\gamma_2/\gamma_1 = 0.4$ and $-0.4$, $E_x$ has an identical profile. This is expected because $\gamma_2$ only influences the motion of vortices along $\mathbf{J}_a$, thus affecting only $E_y$ (ref.~\ref{subsec:inclusion_of_hall_effect}). On the other hand, we find significant difference in $E_y\left(t\right)$ profiles (fig.~\ref{fig:efield2x1}). It is evident that the difference in Hall behavior between $\gamma_2/\gamma_1 = 0.4$ and $-0.4$ can be attributed almost entirely to the transients starting at $t \approx 1500$. Although spiking behavior results in peaks in the opposite direction, they have negligible contribution to the average. Instead, the gradual transient starting at $t \approx 1500$ leads to different profiles for $\gamma_2/\gamma_1 = 0.4$ and $-0.4$, giving a strong positive and negative contribution to Hall field, respectively. We try to relate this transient with order parameter (fig.~\ref{fig:ordPar3x2}) in order to determine the precise behavior causing such transients. We find that this transient is the induced field resulting from variation in $\mathbf{A}$ during the onset of normal state (fig.~\ref{fig:ordPar3x2}). This onset occurs from the right boundary for $\gamma_2/\gamma_1=0.4$ (as seen in fig.~\ref{fig:ordPar3x2}), and from the left for $\gamma_2/\gamma_1=-0.4$ (fig.~\ref{fig:efield2x1}). For both $\gamma_2/\gamma_1=0.4$ and $-0.4$, this transient leads ultimately to saturation of $E_y$ to the same, small positive value. This is the non-flux-flow, or normal-state contribution to the Hall field. 
	Interestingly, we find that although size $l/\xi=15$ exhibits vortex behavior similar to $l/\xi=20$, we do not find sign-reversal in Hall angle. This is due to the suppressed gradual transient fields in $l/\xi=15$, leading to a much weaker contribution of flux-flow Hall effect. Therefore, it is evident that along with $\gamma_2$ (whose value depends on the electronic structure) and several other parameters, size alters Hall behavior significantly, adding to the reasoning behind the observation of a diverse variety of Hall angle profiles in various materials.

	In summary, we have simulated the anomalous Hall effect using the modified TDGL equations in COMSOL Multiphysics and shown that the solutions provide insights into the precise temporal dynamics of transient fields and vortex behavior that scale with the sample size. We have explored theoretically how features of the anomalous Hall effect evolve with a variation of the linear dimensions when the lengths are only a few times the coherence length. The Hall effect behaviour predicted by these simulations may be probed with advanced experimental techniques that have already been applied to image vortices \cite{nv_prappl, nv_natnano, lillie2020laser, squid_nanolett, squid_natcomm, hall_prb, hall_natmater}. Such studies would be important for understanding finite-size effects in superconductors and their evolution at microscopic lengthscales.

	\section*{Acknowledgments}
	K.S. acknowledges financial support from IITB-IRCC Seed grant number 17IRCCSG009, DST Inspire Faculty Fellowship - DST/INSPIRE/04/2016/002284 and AOARD Grant No. FA2386-19-1-4042. SP acknowledges financial support from IRCC, IIT Bombay (17IRCCSG011) and SERB, DST, India (SRG/2019/001419).
	
	\bibliography{main.bib}

\end{document}